\documentclass[aps,prb,reprint,superscriptaddress,longbibliography]{revtex4-2}

\usepackage{graphicx}
\usepackage{color,amssymb,amsmath}
\usepackage[normalem]{ulem}
\usepackage{hyperref}
\usepackage{lipsum}
\usepackage{physics}
\usepackage{float}

\makeatletter
\newcommand*{\rom}[1]{\expandafter\@slowromancap\romannumeral #1@}
\makeatother

\begin{document}
\title{Direct microwave spectroscopy of Andreev bound states in planar Ge Josephson junctions}

\author{M.~Hinderling}
\altaffiliation{These authors equally contributed to this work.}
\affiliation{IBM Research Europe - Zurich, S\"aumerstrasse 4, 8803 R\"uschlikon, Switzerland}

\author{S.~C.~ten~Kate}
\altaffiliation{These authors equally contributed to this work.}
\affiliation{IBM Research Europe - Zurich, S\"aumerstrasse 4, 8803 R\"uschlikon, Switzerland}
	
\author{M.~Coraiola}
\affiliation{IBM Research Europe - Zurich, S\"aumerstrasse 4, 8803 R\"uschlikon, Switzerland}

\author{D.~Z.~Haxell}
\affiliation{IBM Research Europe - Zurich, S\"aumerstrasse 4, 8803 R\"uschlikon, Switzerland}

\author{M.~Stiefel}
\affiliation{IBM Research Europe - Zurich, S\"aumerstrasse 4, 8803 R\"uschlikon, Switzerland}

\author{M.~Mergenthaler}
\affiliation{IBM Research Europe - Zurich, S\"aumerstrasse 4, 8803 R\"uschlikon, Switzerland}

\author{S.~Paredes}
\affiliation{IBM Research Europe - Zurich, S\"aumerstrasse 4, 8803 R\"uschlikon, Switzerland}

\author{S.W.~Bedell}
\affiliation{IBM Quantum, T.J. Watson Research Center, 1101 Kitchawan Road, Yorktown Heights, New York 10598, USA}

\author{D.~Sabonis}
\email{Deividas.Sabonis@ibm.com}
\affiliation{IBM Research Europe - Zurich, S\"aumerstrasse 4, 8803 R\"uschlikon, Switzerland}

\author{F.~Nichele}
\email{fni@zurich.ibm.com}
\affiliation{IBM Research Europe - Zurich, S\"aumerstrasse 4, 8803 R\"uschlikon, Switzerland}

\date{\today}

\begin{abstract}
We demonstrate microwave measurements of the Andreev bound state (ABS) spectrum in planar Josephson junctions (JJs) defined in Ge high mobility two-dimensional hole gases contacted by superconducting PtSiGe. The JJs and readout circuitry are located on separate chips and inductively coupled via flip-chip bonding. For a device with 350~nm junction length, spectroscopic signatures were consistent with the short-junction limit, with an induced superconducting gap $\Delta^{*}\approx48~\mu$eV and transmission $\tau \approx 0.94$. The interaction between the highest-transmission ABS and the resonator was well described by a Jaynes-Cummings model with a vacuum Rabi splitting of approximately 6~MHz. A device with junction length of 1~$\mu$m showed an ABS spectrum consistent with a long junction model. Time-resolved monitoring of the readout resonator in the dispersive regime revealed gate-voltage tunable junction parity fluctuations on the timescale of seconds. Our work indicates a viable path towards hybrid quantum devices based on planar Ge.
\end{abstract}

\maketitle

\section*{Introduction}
Andreev bound states (ABSs) are discrete energy excitations present at the interface between superconductors and semiconductors~\cite{Andreev1964}. They are responsible for the flow of the supercurrent~\cite{BeenakkerABSFormula,Furusaki1991,BTK1982} and, as such, are essential in describing a wide class of physical phenomena. Prominent examples include gate-tunable JJs~\cite{gate-tunable1,Xiang2006} and qubits~\cite{gatemon1,gatemon2,Casparis2018,PitaVidal2020}, and non-reciprocal phenomena~\cite{Baumgartner2022,Matsuo2023}. Particularly interesting is the possibility to investigate ABSs via microwave spectroscopy, which provides a direct access to ABSs energy spectrum of Al break junctions~\cite{Bretheau2013,Janvier2015}, InAs nanowires~\cite{Hays2018,PhysRevX.9.011010,hays2020continuous,Metzger2022Interactions,Bargerbos2022,Wesdorp2024}, and InAs two-dimensional electron gases~\cite{chidambaram2022microwave,hinderling2023flip}. Within the last decade, isolated ABSs have been employed in new qubit architectures, harnessing their charge~\cite{Desposito2001controlled,Zazunov2003Andreev,lee2014spin,Janvier2015,prada2020andreev,aguado2020perspective} and spin~\cite{AndreevSpinQubit,Padurariu2010theoretical,park2017andreev,PhysRevX.9.011010,hays2021coherent,pita2023direct} character. In addition, time-resolved measurements enabled access the incoherent quasiparticle dynamics of ABSs~\cite{PhysRevB.89.104504,Janvier2015,Hays2018,hays2020continuous,hays2021coherent,urbinareport,Erlandsson2023parity,uilhoorn2021quasiparticle,Bargerbos2022,Wesdorp2023parity,Shabani2023parity,pita2023direct}, with reported parity lifetimes in excess of several hundred microseconds~\cite{pita2023direct}. At the same time, these experiments highlighted short spin dephasing times - presumably associated to the nuclear environment of the host material~\cite{hays2021coherent,pita2023direct} - and the need for advanced qubit control techniques. It is therefore important to develop alternative platforms to III-V materials to further investigate the physics of ABSs and mitigate existing limitations towards quantum computing applications. Two-dimensional hole gases in Ge might constitute such a platform thanks to their high mobility, strong spin-orbit interaction, and possibility to provide a nuclear spin-free environment by isotopic purification~\cite{Scappucci2021}. Furthermore, a planar geometry allows for the realization of complex devices, beyond the possibilities of semiconducting nanowires. Several works demonstrated proximity-induced superconductivity in two-dimensional hole gases, where Al was used as superconductor~\cite{Hendrickx2019,GeAl2,aggarwal2021enhancement}. Recently, it was shown that the alloy platinum germanosilicide (PtSiGe) realizes a transparent superconducting contact to Ge, resulting in a hard induced superconducting gap~\cite{tosato2023hard}. However, electrical characterization of induced superconductivity and ABSs in planar Ge has, so far, been limited to dc techniques. 

In this work, we demonstrate a viable platform for isolating discrete ABSs in Ge quantum wells, probe them via microwave spectroscopy and perform real-time parity measurements. Our devices are based on high-mobility two-dimensional hole gases proximitized by PtSiGe contacts. We defined tunable planar JJs by electrostatic gating, and we characterized them by dc and microwave spectroscopy techniques. For the latter, JJs were incorporated in radio frequency superconducting quantum interference devices, which enabled tunability of the phase drop across the junction. Using flip-chip bonding, the superconducting loops were inductively coupled to Nb coplanar microwave resonators located on a low-loss Si substrate~\cite{zellekens2022microwave,hinderling2023flip,hinderling2023parity}. Electrostatic control allowed to operate devices of various junction lengths in a regime where few ABSs at low energy dominated the microwave response. Single-tone spectroscopy revealed hybridization between ABSs and the resonator mode, consistent with a Jaynes-Cummings model. Direct microwave spectroscopy enabled estimation of the induced superconducting gap $\Delta^{*}$ and the mode transmission $\tau$. Monitoring of the readout resonator, operated in the dispersive regime, enabled the observation of JJ parity fluctuations on a time scale of seconds, providing an estimate of the parity lifetime in Ge-based hybrid systems.

\begin{figure*}
	\includegraphics[width=0.95\textwidth]{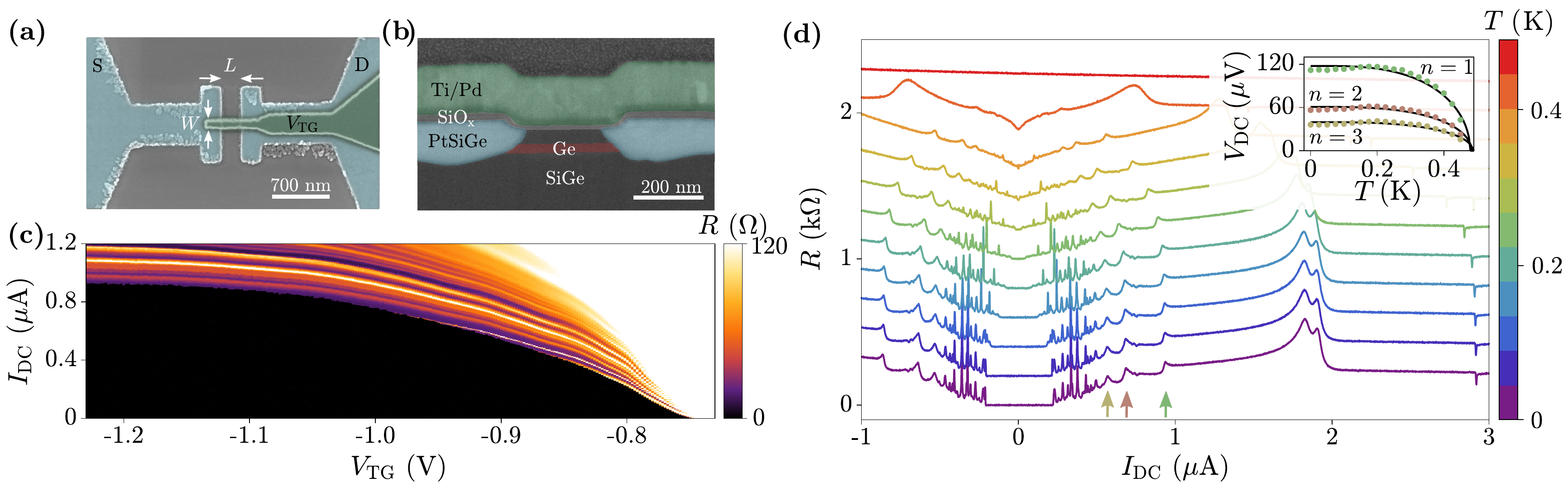}
	\caption{(a) False-colored scanning electron micrograph (top view) of an exemplary Josephson junction (JJ). A conductive channel of length $L$ is formed between the source (S) and the drain (D) PtSiGe electrodes (blue). A Ti/Pd metallic gate of width $W$ (green) is deposited on top of the gate dielectric (not visible). The top-gate voltage labeled $V_\mathrm{TG}$ controls the hole density in the Ge quantum well. (b) Scanning electron micrograph of the cross-section of a typical JJ. The Ge quantum well is colored in red, the superconducting PtSiGe in blue, SiO${}_\mathrm{x}$ dielectric spacer layer in gray and Ti/Pd gate in green. (c) Differential resistance $R$ as a function of bias current $I_\mathrm{DC}$ and top-gate voltage $V_\mathrm{TG}$ for Device~1, which has $L=300$~nm and $W=4~$$\mu$m. The colorscale is saturated. (d) Waterfall plot showing temperature dependence of differential resistance $R$ as a function of current bias $I_\mathrm{DC}$ for multiple values of mixing chamber temperature $T$ (colorbar), where the lowest one is base temperature. The peaks (indicated with arrows) that change position as a function of temperature $T$ are associated to low-order MAR processes. The inset shows the evolution of the three MAR peaks as a function of temperature and voltage drop $V_\mathrm{DC}$ across the JJ, together with fits to the BCS theory (black solid curves).}
	\label{fig1}
\end{figure*}

\section*{Material characterization}

Devices were realized in a strained Ge quantum well located 48~nm below the surface and grown with the remote-plasma chemical vapor deposition technique~\cite{bedell2020low}. The quantum well hosted a high-mobility two-dimensional hole gas with peak mobility $8.3\times10{}^{5}~\mathrm{cm}^2\textrm{(Vs)}^{-1}$ at a carrier density of $2.7\times10{}^{11}~\textrm{cm}^{-2}$, measured in a gated Hall bar. Similar quantum wells were recently employed for realizing high-quality semiconductor spin qubits~\cite{hendrickx2023sweetspot}. Contacts were defined by lift-off of Pt and rapid thermal annealing, resulting in the formation of superconducting PtSiGe. This approach was recently shown to induce a hard superconducting gap in Ge while maintaining the high hole mobility of the system~\cite{tosato2023hard}. A uniform insulating $\mathrm{SiO_{x}}$ layer was deposited via atomic layer deposition, followed by definition of top gates via lift-off of Ti/Pd gate electrodes. A scanning electron micrograph of an exemplary junction (top view) with separation between superconducting contacts $L$ is shown in Fig.~\ref{fig1}(a). A gate of width $W$ was deposited on top of SiO${}_\mathrm{x}$ dielectric layer. The top-gate voltage $V_\mathrm{TG}$ controlled the accumulation of carriers in the junction, which is insulating for $V_\mathrm{TG}=0$. A cross-section image is depicted in Fig.~\ref{fig1}(b), where the quantum well is colored in red, the PtSiGe contacts in blue and the top gate in green. Four devices were studied in this work (Devices~1 to 4), all fabricated following the procedure summarized in Appendix~A.

Device~1 consisted of a planar junction with length $L=300$~nm and width $W=4~\mu$m, connected to four PtSiGe bonding pads for low-frequency electrical measurements. For this, a bias current $I_{\mathrm{DC}}$ was applied to the terminal labeled S in Fig.~\ref{fig1}(a) and the differential resistance $R$ between contacts S and D was measured. This measurement was repeated for multiple $V_{\mathrm{TG}}$ values, with $R$ as a function of $I{_\mathrm{DC}}$ and $V_\mathrm{TG}$ shown in Fig.~\ref{fig1}(c). The switching current $I_{\mathrm{SW}}$, denoting the transition between superconducting and resistive states, was tuned from $0$ up to $930$~nA at $V_\mathrm{TG}=-1.23$~V.
The induced superconducting gap $\Delta^{*}$ was estimated by studying multiple Andreev reflection (MAR) processes. Figure~\ref{fig1}(d) depicts the differential resistance $R$ as a function of bias current $I_\mathrm{DC}$ at multiple mixing chamber temperatures $T$ (see color scale). Resistance peaks marked with arrows are associated to the lowest-order MAR processes. Additional features visible in Fig.~\ref{fig1}(d) for large bias currents, which were also observed in devices based on InAs/Al~\cite{Kjaergaard2017transparent}, are likely related to processes taking place in the superconducting leads of our devices, for example to portions of PtSiGe turning resistive. The inset shows the evolution of the three lowest-order MAR processes as a function of voltage drop $V_\mathrm{DC}$ across the junction, as a function of mixing chamber temperature. The resistance peaks merged at $V_\mathrm{DC}=0$ for $T\approx500$~mK, which serves as an estimate for the critical temperature. The induced superconducting gap $\Delta^{*}$ is obtained from the relation $eV_\mathrm{DC} = 2\Delta^{*}/n$, where $n$ is a positive integer, consistently yielding $\Delta^{*}(0) \approx 60~\mu$eV at base temperature. The MAR position in $V_\mathrm{DC}$ (colored dots) is in agreement with the Bardeen-Cooper-Schrieffer (BCS) relation~\cite{waldram2017superconductivity} $\Delta^{*}(T)=\Delta^{*}(0) \times $tanh(1.74 $\sqrt{T/T_\mathrm{C}-1})$, producing the fits shown as black solid curves for $n=$1, 2, 3 with $\Delta(0)^{*}\approx60$~$\mu$eV and $T_\mathrm{C}\approx480$~mK, consistent with Ref.~\onlinecite{tosato2023hard}. These properties result in a superconducting coherence length in Ge of $\xi_{\mathrm{Ge}}=\hbar v_{\mathrm{F}}/\Delta^{*}=1.8~\mathrm{\mu m}$ at the highest mobility, where $v_{\mathrm{F}}=1.7\times10^5~\mathrm{ms^{-1}}$ is the Fermi velocity.

\begin{figure}[h!]
	\includegraphics[width=0.45\textwidth]{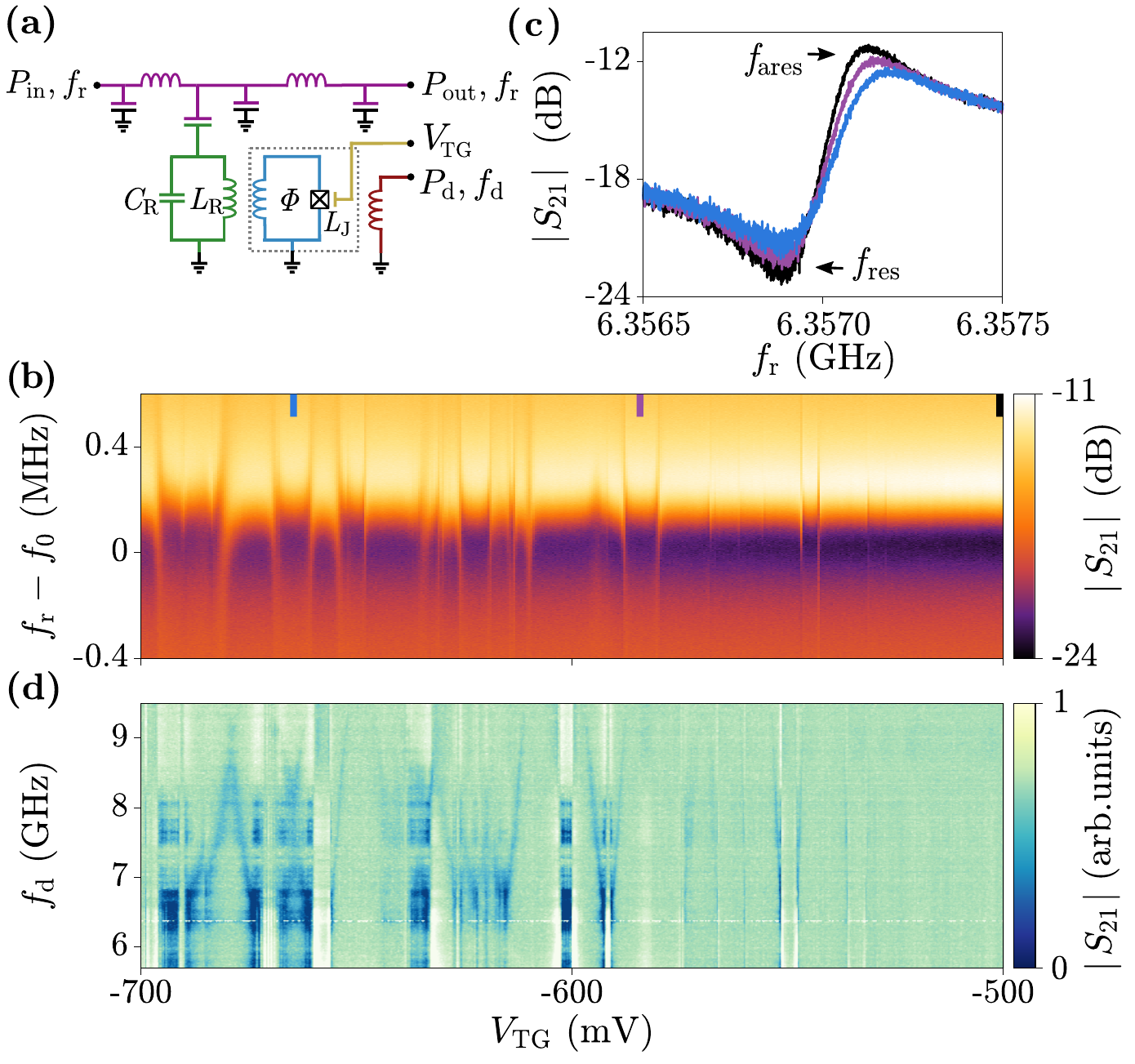}
	\caption{(a) Equivalent circuit of the Ge device, the resonator and the measurement feedline. The Nb resonator (green) and drive line (brown) are placed on a low-loss Si substrate. The PtSiGe superconducting quantum interference device (blue) is inductively coupled to the resonator and the drive line. The low-frequency line (yellow) is used to apply the voltage $V_\mathrm{TG}$, which controls the carrier density in the junction. The resonator is probed in transmission mode through a capacitively coupled feedline (purple). (b) Magnitude of resonator transmission $S_{21}$ as a function of offset readout frequency $f_\mathrm{r}-f_0$ ($f_\mathrm{0}=6.35685$~GHz) and top-gate voltage $V_\mathrm{TG}$ measured at flux $\it{\Phi}=\it{\Phi}$$_0$/2 measured in Device~2. Multiple resonator anticrossings indicate the presence of high transmission Andreev bound states present in the junction. (c) Line cuts of $\abs{S_{21}}$ at selected values of $V_\mathrm{TG}$ [colored markers in (b)] showing the resonance dip ($f_\mathrm{res}$) and antiresonance peak ($f_\mathrm{ares}$) of the resonator response. (d) Magnitude $\abs{S_{21}}$ as a function of the drive tone frequency $f_\mathrm{d}$ ($P_\mathrm{d}=-10$~dBm) for multiple values of $V_\mathrm{TG}$, for the same flux and gate range used for (b). The ABS transition frequency fluctuates as a function of $V_\mathrm{TG}$, resulting in the anticrossings shown in (b) when $f_\mathrm{res}=f_\mathrm{ABS}$. The line-averaged and frequency-dependent background at $V_\mathrm{TG}=-0.5$~V subtracted from the data.}
	\label{fig2}
\end{figure}

\section*{Resonator characterization and measurement circuitry}

Microwave measurements were performed on Device~2 to 4, each consisting of a JJ incorporated into a superconducting PtSiGe loop with an area of $360~\mathrm{\mu m^2}$. A schematic of the measurement circuitry is shown in Fig.~\ref{fig2}(a). The superconducting loop (blue) was inductively coupled to a Nb readout resonator (green), fabricated on a Si chip and capacitively coupled to a coplanar transmission line (purple). An additional line on the Si chip (brown) was used to drive ABS transitions. The device and resonator chips were flip-chip bonded~\cite{zellekens2022microwave,hinderling2023flip,hinderling2023parity}, aligning the loop of each device with the grounded end of a coplanar waveguide resonator, leaving a vacuum gap of approximately $5~\mathrm{\mu m}$. An in-depth description of the resonator chip is found in Ref.~\onlinecite{hinderling2023flip}. Characterization of the superconducting resonators after flip-chip bonding yielded internal (loaded) quality factors of $39000$ ($35000$), $63000$ ($25000$) and $105000$ ($43000$) for Devices~2, 3 and 4, respectively. For tuning the phase drop across the junction and thus controlling the ABS energy, a magnetic flux $\it{\Phi}$ was applied using a home-built superconducting coil mounted on the circuit board that hosted the sample. The data presented in Figs.~\ref{fig2}, ~\ref{fig3} and ~\ref{fig4} were measured using a vector network analyser (VNA). The VNA applied a continuous microwave signal with power $P_\mathrm{in}$ and frequency $f_\mathrm{r}$, close to the resonator resonance at $f_\mathrm{res}$, to the coplanar transmission line and detected the complex scattering parameter $S_\mathrm{21}$ with power $P_\mathrm{out}$. For two-tone spectroscopy measurements, a second tone with source power $P_\mathrm{d}$ and frequency $f_\mathrm{d}$ was applied via a drive line located on the resonator chip. In this case, whenever the gate voltage $V_\mathrm{TG}$ or flux $\it{\Phi}$ was changed, the readout frequency $f_\mathrm{r}$ was adjusted to stay close to the steepest part of the $S_\mathrm{21}$($f_\mathrm{r}$) response (see Appendix~C). For the measurements presented in Fig.~\ref{fig5}, the signal was acquired using a Zurich Instruments SHFQA. For parity detection, the output signal was recorded by digital homodyne mixing, recovering the in-phase ($I$) and quadrature ($Q$) components of the resonator response. More details on the measurement setup are reported in Appendix~B.

We first present microwave characterization of Device~2, with a JJ length $L=350$~nm, which is considerably shorter than $\xi_{\mathrm{Ge}}$, and a resonator bare frequency of $f_\mathrm{0}=6.35685$~GHz. The magnitude of the resonator transmission $S_{21}$ as a function of offset readout frequency $f_\mathrm{r}-f_\mathrm{0}$ and top-gate voltage $V_\mathrm{TG}$ is shown in Fig.~\ref{fig2}(b). At $V_\mathrm{TG}=-0.5$~V, the junction was depleted and the resonator approached $f_\mathrm{0}$. Setting $V_\mathrm{TG}$ more negative, holes were accumulated in the channel and the resonator frequency experienced a repeated pattern of vacuum Rabi splittings with multiple energy levels present in the junction. At each position of vacuum Rabi splitting, the resonator interacted with one or more ABSs in the junction, satisfying the resonance condition $f_\mathrm{res}$ = $f_\mathrm{ABS}=2E_\mathrm{ABS}/h$. The large number of resonator anti-crossings in Fig.~\ref{fig2}(b) suggests that high-transmission modes are ubiquitous in the JJ. 

Figure~\ref{fig2}(c) shows the magnitude of $S_{21}$ as a function of readout frequency $f_\mathrm{r}$ recorded at selected $V_\mathrm{TG}$ values [colored markers in Fig.~\ref{fig2}(b)]. The frequency of the minimum in $\abs{S_{21}}$, indicated as $f_\mathrm{res}$, was independent of $V_\mathrm{TG}$ and was close to the bare frequency of the resonator. On the contrary, the maximum (antiresonance feature) located around $f_\mathrm{ares}=6.35713$~GHz shifted to higher frequencies as $V_\mathrm{TG}$ was reduced. This frequency shift is attributed to a change in the inductance of the junction, which depends on the density of charge carriers and the ABS spectrum in the JJ. The asymmetry in $\abs{S_{21}}$ with respect to its minimum was likely caused by a combination of direct leakage of the readout signal between the input and output ports of the feedline, impedance mismatches in the signal propagation path, and dielectric losses in the system. Figure~\ref{fig2}(d) shows a direct microwave spectroscopy measurement of the junction acquired at $\it{\Phi}=\it{\Phi}$$_0$/2, where $\it{\Phi}$$_0$ is the superconducting flux quantum, and spanning the same $V_\mathrm{TG}$ range as Fig.~\ref{fig2}(b). The ABS transition frequency fluctuated as a function of $V_\mathrm{TG}$ and crossed the resonator frequency multiple times, giving rise to the vacuum Rabi splittings observed in Fig.~\ref{fig2}(b). We note that features in Figs.~\ref{fig2}(b) and (d) are shifted with each other by up to 10~mV. The shift is caused by gate-voltage-induced hysteresis, commonly observed in SiGe-based devices when gate voltages are swept over large ranges. This effect was thoroughly discussed in a recent work that employed quantum wells similar to the one presented here~\cite{massai2023impact}. 

\section*{Microwave spectroscopy in a short junction}
\begin{figure}[h]
	\includegraphics[width=0.45\textwidth]{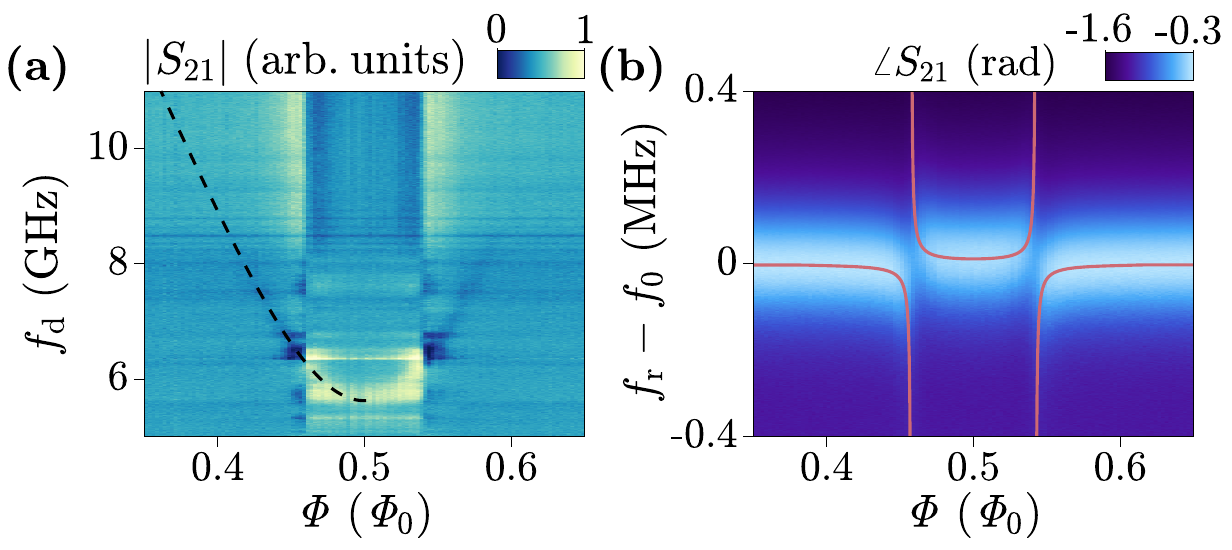}
	\caption{(a) Magnitude of resonator transmission $S_{21}$ as a function of the drive tone frequency $f_\mathrm{d}$ and external flux $\it{\Phi}$ recorded at $V_\mathrm{TG}=-0.5675$~V in Device~2. A fit of the ABS energy profile using the short junction model is plotted partially on top (black dashed line), yielding an induced superconducting gap $\Delta^{*} \approx 48~\mu$eV and transmission $\tau \approx 0.94$. (b) Phase of resonator transmission $S_{21}$ measured simultaneously with (a), as a function of offset readout frequency $f_\mathrm{r}-f_\mathrm{0}$. The prediction based on a Jaynes-Cummings model is plotted on top (orange). The ABS and resonator interact at $\it{\Phi}$ = 0.46$\it{\Phi}$$_0$ and $\it{\Phi}$ = 0.54$\it{\Phi}$$_0$, where $f_\mathrm{res}$ = $f_\mathrm{ABS}$. The best agreement between the experimental data and the model is obtained assuming an ABS-resonator coupling of $g/2\pi \approx$ 3~MHz.}
	\label{fig3}
\end{figure}

Next, we proceed with detailed spectroscopy of an ABS in Device~2. For this, we set $V_\mathrm{TG}=-0.5675$~V and recorded the dependence of the magnitude of the resonator transmission $S_{21}$ on a second drive tone frequency $f_\mathrm{d}$ and on the external flux $\it{\Phi}$. The resulting ABS energy profile is shown in Fig.~\ref{fig3}(a). Here, the clearest spectral lines were obtained around $\it{\Phi}=\it{\Phi}$$_0$/2, where the coupling between the device and the resonator was strongest~\cite{Janvier2015}. The ABS excitation frequencies were extracted and fitted to the dispersion relation valid in a short JJ:
\begin{equation}
	E_\mathrm{ABS}=\pm \Delta^{*} \sqrt{1-\tau \sin^2 (\mathrm{\pi} \it{\Phi}/ \it{\Phi}_\mathrm{0})},
	\label{eq1}
\end{equation}
where $\tau$ is the transmission of the ABS~\cite{BeenakkerABSFormula,BagwellABSFormula} and $\Delta^{*}$ is the induced superconducting gap in the leads. The fit is plotted in the left half of Fig.~\ref{fig3}(a) as a black dashed curve on top of the experimental data. The best fit produced the values for induced superconducting gap $\Delta^{*}=48~\mu$eV and the transmission $\tau=0.94$. The vertical lines at $\it{\Phi}$ = 0.46$\it{\Phi}$$_0$ and $\it{\Phi}$ = 0.54$\it{\Phi}$$_0$ are attributed to shifts in the resonance frequency $f_\mathrm{res}$, caused by the interaction between the resonator and the ABS at $f_\mathrm{res}$ = $f_\mathrm{ABS}=2E_\mathrm{ABS}/h$. The value of $\Delta^{*}$ deduced from microwave spectroscopy measurements differs by 20\% from that obtained from MAR studies presented in Fig,~\ref{fig1}. The cause of this discrepancy is not understood. A possibility is that the ABS in Fig.~\ref{fig3} was affected by additional ABSs lying at higher frequencies~\cite{Metzger2022Interactions}, outside of the measurement window of our setup, making the treatment with Eq.~\ref{eq1} not fully accurate. Furthermore, the difference between the two superconducting gaps might originate from trivial sample-to-sample variations.

Figure~\ref{fig3}(b) depicts the phase of the resonator transmission $S_{21}$ recorded simultaneously with the data in Fig.~\ref{fig3}(a) as a function of offset readout frequency $f_\mathrm{r}-f_\mathrm{0}$ and flux $\it{\Phi}$. Vacuum Rabi splittings were observed when the resonator interacted with the ABS, satisfying the resonance condition $f_\mathrm{res}$ = $f_\mathrm{ABS}$ ($\it{\Phi}$ = 0.46$\it{\Phi}$$_0$ and $\it{\Phi}$ = 0.54$\it{\Phi}$$_0$). The resonant interaction between the single mode of the microwave resonator and the discrete ABS is described by a Jaynes-Cummings model, producing hybridized ABS-resonator states with frequency:
\begin{equation*}
\begin{split}
	\omega'_\mathrm{ABS-res}&=\frac{1}{\sqrt{2}} \biggl[ \omega^{2}_\mathrm{ABS}+\omega^{2}_\mathrm{res} \biggr. \\
& \biggl. \pm \sqrt{16g^2 \omega^{}_\mathrm{ABS} \omega_\mathrm{res}+(\omega^{2}_\mathrm{ABS}-\omega^{2}_\mathrm{res})^{2}} \biggr] ^{1/2},
\end{split}
	\label{eq2}
\end{equation*}
where $\omega^{}_\mathrm{ABS}=2\pi f_\mathrm{ABS}$, $\omega_\mathrm{res}=2\pi f_\mathrm{res}$ and $g$ is the coupling constant characterizing the strength of the interaction~\cite{Upadhyay2021coupling}. The calculated resonator response in the presence of vacuum Rabi splitting is plotted in orange on top of the experimental data, using the ABS energy profile from Eq.~\ref{eq1} and bare resonator frequency as input parameters. The closest agreement with the experimental data was achieved with $g/2\pi \approx 3$~MHz.

\section*{Microwave spectroscopy in a long junction}

\begin{figure}[h]
	\includegraphics[width=0.45\textwidth]{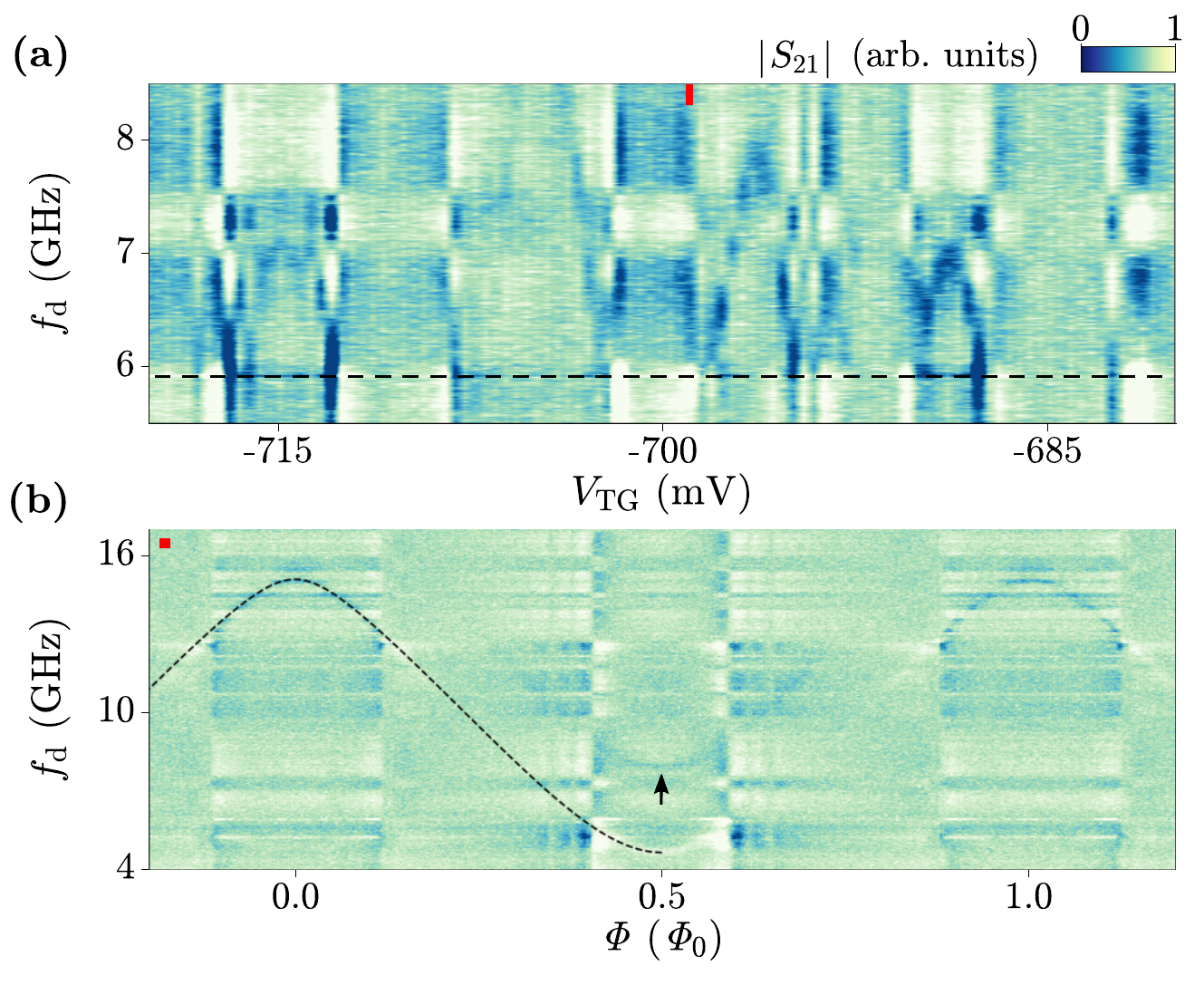}
	\caption{(a) Magnitude of resonator transmission $S_{21}$ as a function of the drive tone frequency $f_\mathrm{d}$ ($P_\mathrm{d}=5$~dBm) and gate voltage $V_\mathrm{TG}$, measured at flux $\it{\Phi}=\it{\Phi}$$_0$/2 in Device~3. The resonance frequency of the resonator is indicated by a dashed line. (b) Magnitude of resonator transmission $S_{21}$ as a function of the drive tone frequency $f_\mathrm{d}$ ($P_\mathrm{d}=1.5$~dBm) and flux $\it{\Phi}$ at $V_\mathrm{TG}=-0.699$~V [red marker in (a)], together with a pair transition of the long-junction model (black dashed line). A second ABS transition is indicated with a black arrow.}
	\label{fig4}
\end{figure}

Device~3 had design and readout parameters similar to Device~2, except for a junction length of $L=1$~$\mu$m, which is approximately half of $\xi_{\mathrm{Ge}}$. The magnitude of the resonator transmission $S_{21}$ as a function of a second drive tone frequency $f_\mathrm{d}$ ($P_\mathrm{d}=5$~dBm) and gate voltage $V_\mathrm{TG}$ is shown in Fig.~\ref{fig4}(a). The ABS transition frequency varied as a function of $V_\mathrm{TG}$ and crossed the resonator frequency (dashed line) several times. In Fig.~\ref{fig4}(b), the magnitude of the resonator transmission $S_{21}$ is depicted as a function of the drive tone frequency $f_\mathrm{d}$ ($P_\mathrm{d}=1.5$~dBm) and flux $\it{\Phi}$ at $V_\mathrm{TG}=-0.699$~V [red marker in Fig.~\ref{fig4}(a)]. The minimum ABS transition frequency at $\it{\Phi}=\it{\Phi}_\mathrm{0}/\mathrm{2}$ was lower than the ABS transition frequency at $V_\mathrm{TG}=-0.699$~V in Fig.~\ref{fig4}(a). This is attributed to the gate-voltage-induced hysteresis mentioned above. At $\it{\Phi}$~$=0$, the ABS transition frequency was $f_\mathrm{ABS}\approx 15$~GHz~$<2\Delta^{*}/h$. This implies that the JJ was not in the short junction limit and thus required theoretical treatment beyond Eq.~\ref{eq1}. For a junction with $L \approx \xi_{\mathrm{Ge}}$, the spin-degenerate ABS energy, $E_\mathrm{ABS}$ is given by the solution of the transcendental equation
\begin{equation}
\begin{split}
	\tau \cos(\varphi) + (1-\tau) \cos(2\lambda \epsilon x_\mathrm{r}) = \cos(2 \arccos(\epsilon) - 2\lambda \epsilon),
\end{split}
	\label{eq3}
\end{equation}
where $\lambda = L/\xi_{\mathrm{Ge}}$, $\epsilon=E_\mathrm{ABS}/\Delta^{*}$, $\varphi$ is the phase difference across the JJ and $x_\mathrm{r}=2x_0/L$ is a model parameter given by the position $x_0 \in [-L/2,L/2]$ of the impurity characterized with transmission $\tau$~\cite{PhysRevX.9.011010}. The black dashed line, plotted on top of the experimental data in the left half of Fig.~\ref{fig4}(b), is a fit to Eq.~\ref{eq3} for $\lambda=0.87$, $\tau=0.835$ and $x_\mathrm{r}=0.6$. The $\lambda$ derived from this analysis is consistent with the value of $\xi_{\mathrm{Ge}}$ deduced from the MAR analysis of Fig.~\ref{fig1}(d). Figure~\ref{fig4}(b) shows an additional spectral line, likely associated with another pair transition (see black arrow). As this feature was particularly sharp and located far from circuit resonances, it was employed to estimate the inhomogeneous dephasing time $T^*_\mathrm{2}$ in this device. Our analysis, reported in Appendix~D, yielded $T^*_\mathrm{2}\sim 1~\mathrm{ns}$. We did not observe evidence of spin-split ABSs, which have been detected with microwave spectroscopy in long InAs junctions~\cite{PhysRevX.9.011010,hays2020continuous,Metzger2022Interactions,Wesdorp2024}. It is unclear why transitions of spin-split ABSs were not observed, and is an open question to be addressed in future work. This could be done by systematically studying JJs with different lengths, possibly extending to values longer than $1~\mathrm{\mu m}$.

\section*{Detection of parity fluctuations of Andreev level}

Finally, we present real-time detection of parity fluctuations of ABSs. The parity detection scheme was based on the continuous monitoring of the resonator transmission at frequency $f_\mathrm{r}$, which was affected by changes in the JJ inductance resulting from the trapping and release of an unpaired quasiparticle. This measurement technique requires a sufficient coupling strength $g$ between the ABS and the resonator, large $Q$ factor for an enhanced detection sensitivity and low levels of noise in the system. The microwave signal amplification chain used for the measurements in Figs.~\ref{fig2},~\ref{fig3} and~\ref{fig4} did not provide a sufficient signal-to-noise ratio (SNR) for time-resolved detection. Therefore, the setup was upgraded with an in-house built traveling wave parametric amplifier (TWPA) installed at the mixing chamber plate, providing up to 20 dB of amplification in the frequency range of $4-8$~GHz (see Appendix~B).

\begin{figure}[h]
	\includegraphics[width=0.45\textwidth]{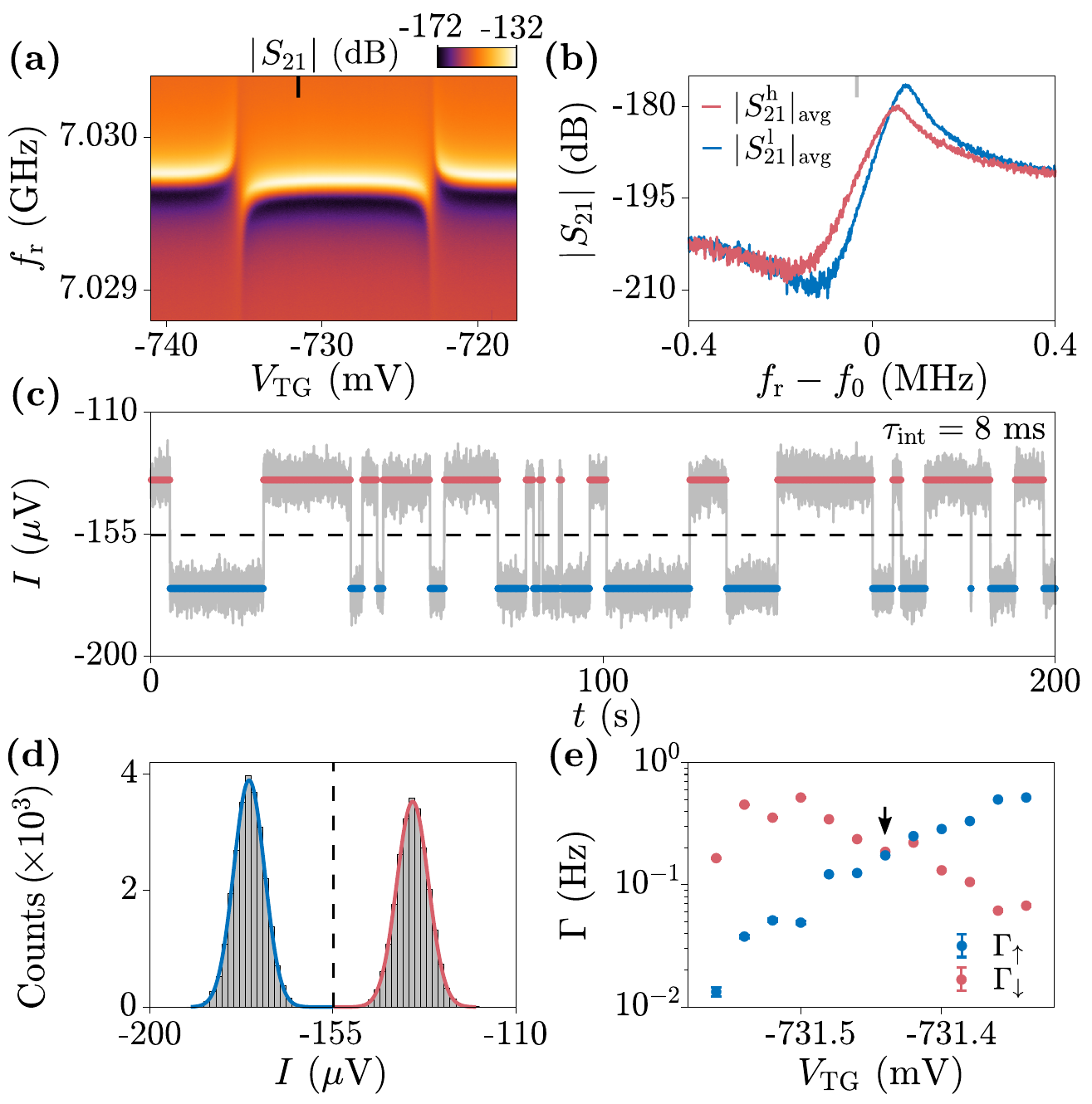}
	\caption{(a) Magnitude of resonator transmission $S_{21}$ as a function of readout frequency $f_\mathrm{r}$ and top-gate voltage $V_\mathrm{TG}$ measured in Device~4, showing two ABS--resonator anticrossings. (b) Averaged ($N$ = 19) $\abs{S_{21}}$ response as a function of offset readout frequency $f_\mathrm{r}-f_0$, where red and blue traces correspond to two junction parity states, labeled as $\abs{S_{21}^\mathrm{h}}$$_\mathrm{avg}$ and $\abs{S_{21}^\mathrm{l}}$$_\mathrm{avg}$ associated with high (h) and low (l) resonator transmission, respectively. (c) A section of the time-resolved resonator response as a function of measurement time $t$ (gray data points, $V_\mathrm{TG}=-731.44$~mV, $\tau_\mathrm{int}=8$~ms) measured at the readout frequency $f_\mathrm{r}$ = 7.02964~GHz, indicated with the grey marker in (b). The data is plotted along the maximum visibility axis of the in-phase quadrature ($I$). Red and blue lines indicate the two states and the black dashed line indicates the threshold. (d) Histogram distribution of the resonator response in (c) together with a double-gaussian fit, used for the binning assignment. (e) Parity switching rates $\Gamma_\mathrm{\uparrow}$ and $\Gamma_\mathrm{\downarrow}$ as a function of $V_\mathrm{TG}$ in the vicinity of the black marker in~(a). The black arrow indicates the gate setting at which the time trace depicted in (c) was recorded.}
	\label{fig5}
\end{figure}

The parity measurement was performed in Device~4, which was lithographically identical to Device~2 but showed larger anticrossings, indicative of a stronger resonator--ABS interaction, and a higher resonator quality factor. Additionally, to increase the number of switching events, the measurement was performed at a mixing chamber temperature of 80~mK. The magnitude of the resonator transmission $S_{21}$ as a function of readout frequency $f_\mathrm{r}$ is shown in Fig.~\ref{fig5}(a) for a $V_\mathrm{TG}$ range containing two large anticrossings. Figure~\ref{fig5}(b) depicts two frequency responses $\abs{S_{21}}$, both obtained for $V_\mathrm{TG}\approx-731~\mathrm{mV}$, but at selected points in time. We associate the two resonator responses, characterized by a difference in frequency shift and damping, to two parity states and refer to them as $\abs{S_{21}^\mathrm{h}}$$_\mathrm{avg}$ and $\abs{S_{21}^\mathrm{l}}$$_\mathrm{avg}$ (see Appendix~F for details). By fixing the readout frequency using resonator compensation, time-resolved resonator responses were recorded for 400~s. Part of an exemplary trace is shown in Fig.~\ref{fig5}(c), where transitions between two levels of the in-phase component $I$ were detected. The red and blue lines indicate state assignment using a two-state model, and the dashed line at $I=-155$~$\mu$eV shows the threshold used to discriminate the parity states. A more elaborate description of parity detection and state assignment is reported in Appendices~E and~F. The average rates for parity switching $\Gamma_\mathrm{\uparrow}$ and $\Gamma_\mathrm{\downarrow}$ were estimated by counting the number of transitions in the acquisition window and dividing by the total time spent in each state. The resonator response shown in Fig.~\ref{fig5}(c) was characterized by $\Gamma_\mathrm{\uparrow} = 0.174 \pm 0.002$~Hz and $\Gamma_\mathrm{\downarrow} = 0.186 \pm 0.002$~Hz. The same procedure was repeated as a function of $V_\mathrm{TG}$ with 20~$\mu$V step size at $\it{\Phi}$ = $\it{\Phi}$$_0/2$. The results are summarized in Fig.~\ref{fig5}(e), with red and blue data points corresponding to $\Gamma_\mathrm{\downarrow}$ and $\Gamma_\mathrm{\uparrow}$, respectively. Both rates show a dependence on $V_\mathrm{TG}$, with a rate inversion at $V_\mathrm{TG}\approx-731.44$~mV. Parity switching times obtained with this procedure are very long, orders of magnitude longer than parity lifetimes reported for Al break junctions~\cite{Janvier2015} and InAs nanowires~\cite{Hays2018,Bargerbos2022}. Such result could be indicative of a particularly clean quasiparticle environment in Ge/PtSiGe quantum wells, good shielding from external perturbations in our measurement setup, or protection from quasiparticle poisoning provided by residual charging energy in the JJ. The inversion of the parity switching rates in Fig.~\ref{fig5}(e) is suggestive of Coulomb blockade~\cite{urbinareport,hinderling2023parity}, which could be caused by disorder or non-uniform gating in the Ge quantum well. For instance, the electric field generated by the top-gate electrode could be partially screened close to the PtSiGe contacts [see Fig.~\ref{fig1}(b)], resulting in a locally reduced hole density. Extensive investigation of parity fluctuations across multiple devices would help clarify the origin and reproducibility of the parity lifetime behavior reported here.

\section*{Conclusion}\label{sec13}

In summary, we have performed electrical measurement of hybrid JJs in a high-mobility two-dimensional hole gas in Ge, where the quantum well was proximitized by superconducting PtSiGe contacts. Investigation of MAR processes with dc techniques yielded a critical temperature $T_\mathrm{C}\approx480$~mK and an induced superconducting gap $\Delta^{*}\approx60$~$\mu$eV. Coupling to microwave resonators was enabled by embedding the JJs in a PtSiGe loop and flip-chip bonding. Two-tone spectroscopy revealed that a 350~nm long JJ hosted discrete ABSs, which could be described by a short junction model with transmissions above 0.94 and $\Delta^{*}\approx48$~$\mu$eV. The spectroscopic signatures conformed to a Jaynes-Cummings model with $g/2\pi=3$~MHz. Microwave spectroscopy in a junction with length 1~$\mu$m was instead better described by a long-junction model. In both cases, the ABSs residing in the junction showed hybridization with the resonator.  Time-resolved monitoring of the resonator in the dispersive regime enabled the detection of parity fluctuations in a short junction. The time scales characterizing parity stability were of the order of several seconds and could be controlled electrostatically, resulting in long parity lifetimes. Detailed studies of parity lifetimes in planar Ge JJs and the role of residual charging energies are needed for gaining a better understanding of the dynamics governing parity fluctuations in these devices and laying the foundation for high-coherence hybrid qubits in planar Ge.

In the light of our results, two-dimensional hole gases in Ge constitute a promising platform for future investigation of ABS physics and realization of hybrid quantum devices. We have shown that PtSiGe contacts, while providing a smaller induced gap than Al, allow for a simple realization of high-quality devices hosting isolated ABSs, both in the short- and long-junction regime. We have also demonstrated that the flip-chip approach enables the integration of coplanar microwave resonators with hybrid Ge devices, with large quality factors and coupling coefficients, allowing for time-resolved parity detection. Such insights will guide towards a new generation of quantum devices, taking advantage of the high mobility, strong spin-orbit interaction and nuclear-spin free environment of Ge two-dimensional hole gases.

\section*{APPENDIX A: DEVICE FABRICATION}
\begin{figure*}
	\includegraphics[width=0.95\textwidth]{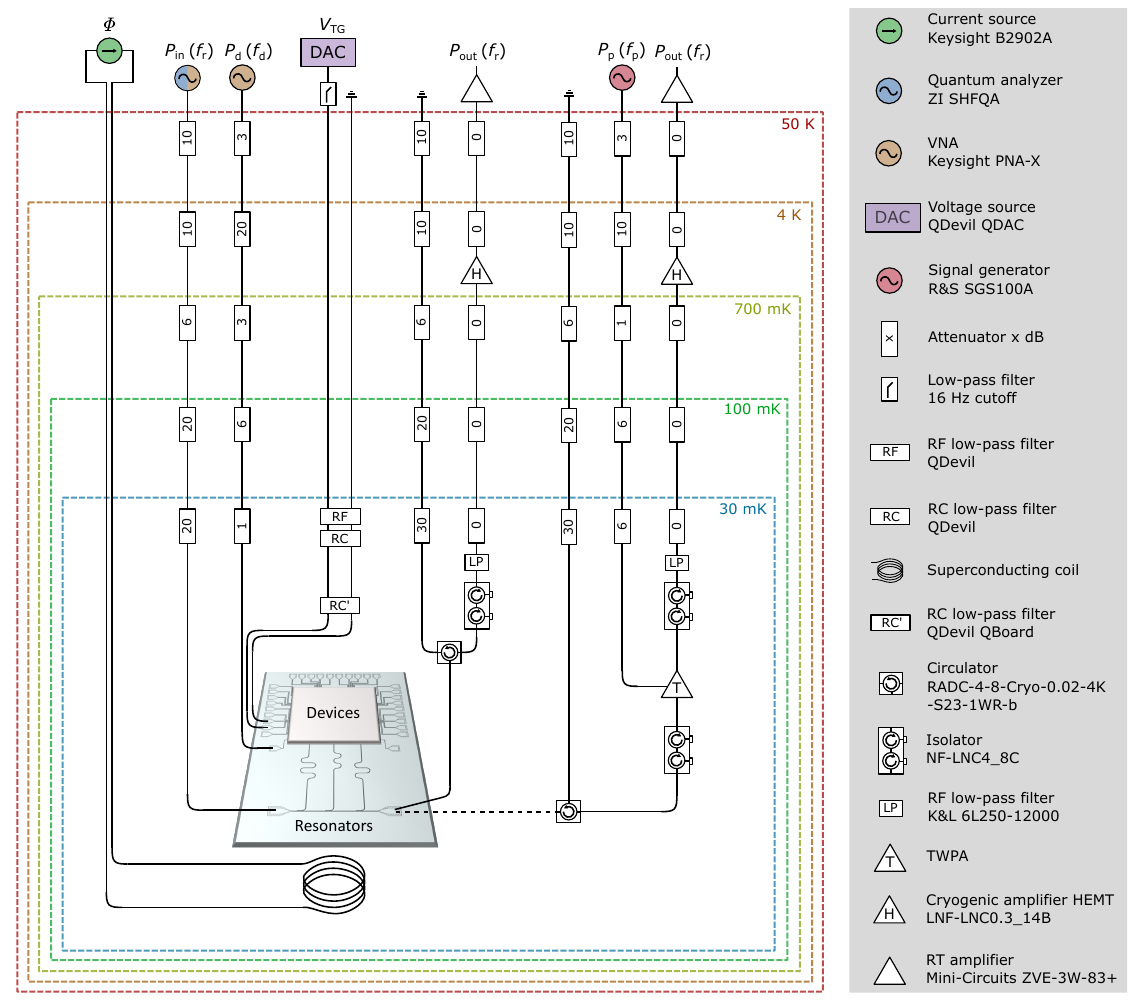}
	\caption{Schematic representation of the flip-chip bonded device, the dilution refrigerator used to perform the measurements and the electrical setup.}
	\label{fig6}
\end{figure*}

The superconductor-semiconductor planar Josephson junction devices were fabricated from a Ge/SiGe heterostructure grown in reduced-pressure chemical vapor deposition reactor~\cite{bedell2020low}. The stack consisted of a strain-relaxed Ge layer on a Si wafer and a SiGe buffer with increasing Si content to reach the $\mathrm{Si_{0.2}Ge_{0.8}}$ stoichiometry used for the quantum well barriers. The quantum well was a 20~nm thick layer of Ge located 48~nm below the surface of the wafer. Electrical contacts were obtained by electron beam lithography, evaporation and lift-off of Pt. A few second dip in buffered HF was performed prior to evaporation to remove native oxides. Following evaporation, Pt was diffused in the heterostructure in a rapid thermal annealer at a temperature of 350° C, resulting in germanosilicide (PtSiGe). A 12~nm thick layer of $\mathrm{SiO_{x}}$ was deposited over the entire sample using atomic layer deposition, before evaporating Ti/Pd (5~nm/80~nm) gate electrodes. Finally, the chip was diced into a 3~mm by 3~mm pieces and flip-chip bonded. The fabrication process of the resonators and the flip-chip bonding procedure are explained in detail in Refs.~\cite{hinderling2023flip} and \cite{hinderling2023parity}.

\section*{APPENDIX B: MEASUREMENT SETUP}
Measurements were performed in a BlueFors BF~LD~400 cryogen-free dilution refrigerator with a mixing chamber base temperature of 30~mK. Figure~\ref{fig6} illustrates the wiring of the dilution refrigerator. The dc measurements of Fig.~\ref{fig1} were taken with a Stanford Research SR860 lock-in amplifier and home-made voltage amplifiers. The data for Figs.~\ref{fig2} to \ref{fig4} in the Main Text was measured with a Keysight PNA-X vector network analyzer, while the data for Fig.~\ref{fig5} was acquired with a Zurich Instruments SHFQA 8.5 GHZ Quantum Analyzer (in a different cool-down). As shown in Fig.~\ref{fig6}, both instruments were used to apply a readout tone with power $P_\mathrm{in}$ and frequency $f_\mathrm{r}$. Two-tone spectroscopy measurements were performed by applying a continuous drive tone with power $P_\mathrm{d}$ and frequency $f_\mathrm{d}$. The readout and drive signals were attenuated by 66~dB and 33~dB, respectively. For the measurements performed with the VNA, the signal transmitted through the readout transmission line passed through a circulator, a dual isolator and an rf low-pass filter before amplification. A different rf readout line, with a traveling wave parametric amplifier (TWPA) with 20~dB gain, was used for the measurements in Fig.~\ref{fig5} of the Main Text. Both rf readout lines were equipped with a cryogenic high electron mobility transistor (HEMT) amplifier, installed at the 4~K stage, and a room-temperature amplifier. The amplified signal was detected at the port labeled as $P_\mathrm{out}$. A QDevil digital-to-analog converter dc voltage source was used to apply the gate voltage $V_\mathrm{TG}$. The dc line was filtered using a home-made low-pass filter at room temperature, QDevil RF and RC filters at the mixing chamber stage of the dilution refrigerator, and an RC filter integrated in the printed circuit board which carried the flip-chip bonded chips. A home-made superconducting coil, mounted on top of the printed circuit board and current-biased by a Keysight B2902A SMU, provided the magnetic flux threading the device loop. The sample space was shielded from external magnetic fields by a home-made magnetic shield consisting of a mu-metal and a superconducting layer.

\section*{APPENDIX C: RESONATOR COMPENSATION}

The two-tone spectroscopy measurements in Figs.~\ref{fig2}, \ref{fig3} and \ref{fig4}, as well as the time traces shown in Fig.~\ref{fig5}, were acquired using resonator compensation. For each change in gate voltage in Figs.~\ref{fig2}(d), \ref{fig3}(a) and \ref{fig4}(a), or flux in Fig.~\ref{fig4}(b), $\abs{S_{21}}$ was recorded as a function of readout frequency $f_\mathrm{r}$. Then, the resonator response minimum was found and the readout frequency for the two-tone spectroscopy measurement was fixed at 120~kHz offset from this minimum. Before plotting the data, the median of the resonator response along the voltage axis was subtracted from all data points, resulting in maps of compensated resonator response as a function of drive frequency $f_\mathrm{d}$. The time traces used to extract the parity switching rates $\Gamma_{\uparrow}$ and $\Gamma_{\downarrow}$ in Fig.~\ref{fig5}(e) were acquired using resonator compensation with an offset of $-50$~kHz from the resonator response maxima.

\section*{APPENDIX D: LONG JUNCTION ADDITIONAL DATA}

\begin{figure}[b]
	\includegraphics[width=0.45\textwidth]{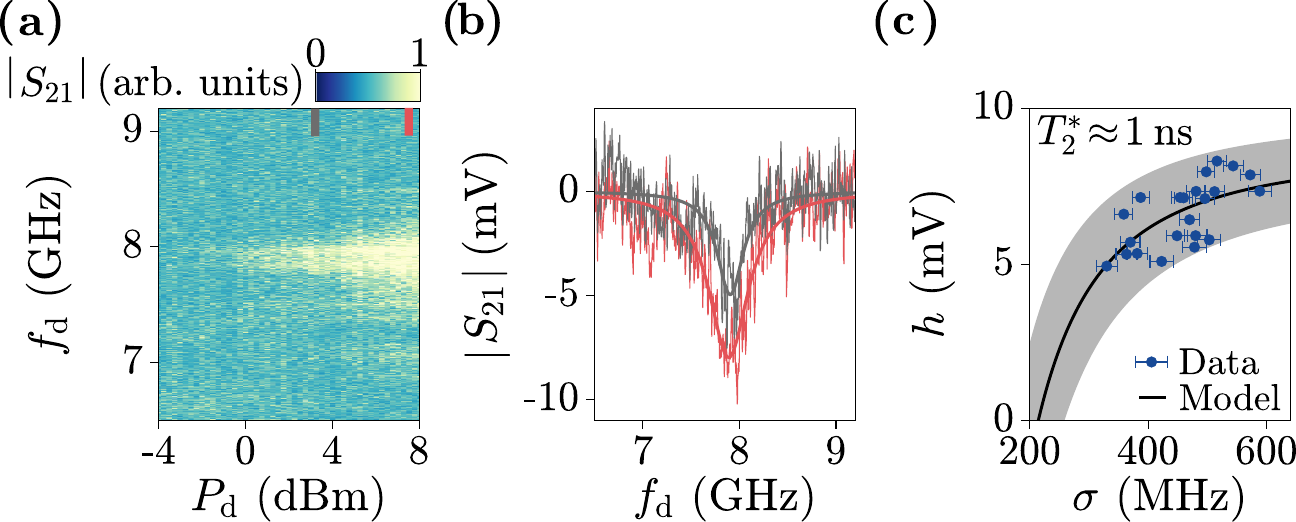}
	\caption{(a) Magnitude of resonator transmission $S_{21}$ as a function of the drive tone frequency $f_\mathrm{d}$ and drive power $P_\mathrm{d}$ at $\it{\Phi}=\it{\Phi}$$_0$/2 and $V_\mathrm{TG}=-0.699$~V. The colorscale is saturated. (b) Line cuts at the marker positions in (a) together with fits to a Lorentzian lineshape model. (c) Parametric plot of spectral line amplitude $h$ as a function of the linewidth $\sigma$ (blue points), together with a fit to Eq.~\ref{eq4} (black solid curve) and the standard deviation of the fit (grey area). The minimum extrapolated linewidth $\sigma_\mathrm{min} \approx 215$~MHz at $h=0$ defines the inhomogeneous dephasing time $T^*_\mathrm{2} \approx 1$~ns of the pair transition.}
	\label{fig7}
\end{figure}

To estimate the inhomogeneous dephasing time $T^*_\mathrm{2}$, the spectroscopic feature indicated with a black arrow in Main Text Fig.~\ref{fig4}(b) was investigated at flux $\it{\Phi}$ = $\it{\Phi}$$_0/2$ as a function of drive power $P_\mathrm{d}$. This ABS transition was studied because it was clearly visible and isolated from standing waves. For every $P_\mathrm{d}$ value in Fig.~\ref{fig7}(a), the spectral linecut was fit to a Lorentzian lineshape profile to extract the spectral line amplitude $h$ and the linewidth $\sigma$. Two spectroscopic lines, together with the fits, are shown in Fig.~\ref{fig7}(b), taken at the marker positions in Fig.~\ref{fig7}(a). The spectral line amplitude $h$ and linewidth $\sigma$ are plotted in Fig.~\ref{fig7}(c) together with a fit assuming 
\begin{equation}
	h=h_\mathrm{0}(1-\sigma_\mathrm{min}^2/\sigma^2),
	\label{eq4}
\end{equation}
where $h_0$ is a prefactor and $\sigma_\mathrm{min}$ is the spectral linewidth at $h=0$~\cite{Petersson2010linewidth}. The best fit with $\sigma_\mathrm{min}=215 \pm 22$~MHz was used to estimate the inhomogeneous dephasing time $T^*_\mathrm{2}=1/(\sqrt{2}\pi \sigma_\mathrm{min}) \approx$~1~ns. This value for $T^*_\mathrm{2}$ is of the same order of magnitude as values previously reported for other condensed matter systems~\cite{Petersson2010linewidth,Shi2013,hinderling2023flip}.

\section*{APPENDIX E: PARITY SWITCHING RATE ESTIMATION}

The parity switching rates in Fig.~\ref{fig5}(e) were estimated using time traces of the in-phase and quadrature components of the resonator response, where each time trace was measured at a different gate voltage. The readout frequency was adjusted using resonator compensation and readout power of $P_\mathrm{in}=-39.1$~dBm, setting resonator in the dispersive regime with an intermediate photon number. For each value of $V_\mathrm{TG}$, 50001 data points were recorded with an integration time of 8~ms, resulting in a 400~s time trace. The upper bound of the detectable parity switching rates is 125~Hz, set by the integration time. The lower bound of the detectable parity switching rates is $2.5 \times 10^{-3}~\mathrm{Hz}$, set by the total measurement time.

\begin{figure}
	\includegraphics[width=0.45\textwidth]{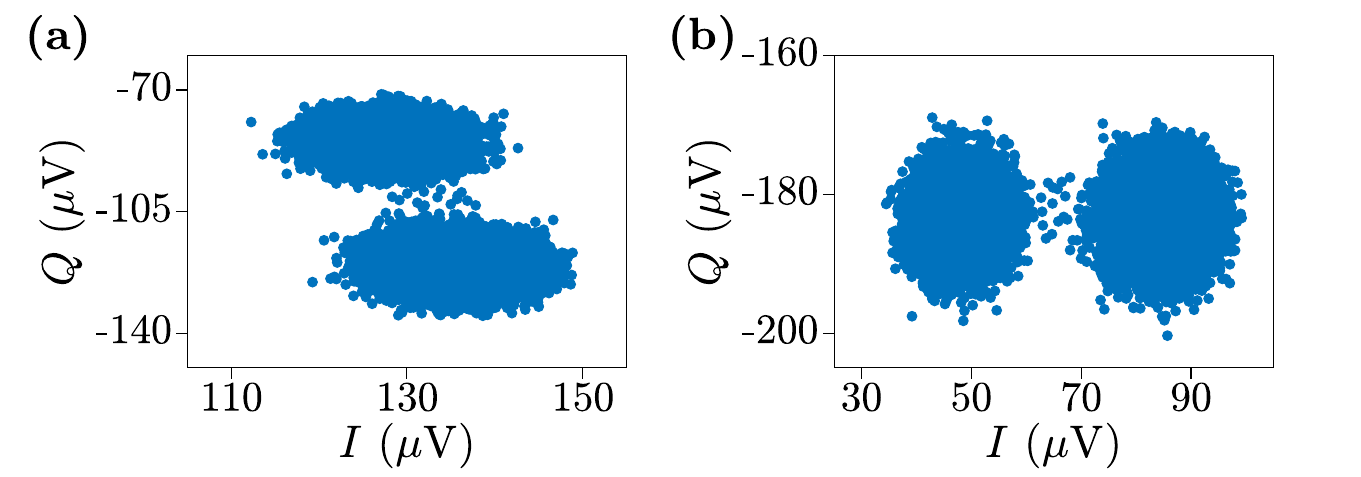}
	\caption{(a) Raw data of the in-phase ($I$) and quadrature ($Q$) components of the complex resonator response. (b) Rotated resonator response. After data rotation, the parity switching information is encoded in in-phase quadrature $I$. The plots were obtained from a time trace of 400~s, measured with 8~ms integration time.}
	\label{fig8}
\end{figure}

The data of each time trace which showed parity switching [e.g. Fig~\ref{fig8}(a)] was rotated such that all information was encoded in the in-phase component of the signal, $I$ [Fig.~\ref{fig8}(b)]. Subsequently, $I$ was converted into a histogram distributed into 50 bins. This histogram was then fitted to the sum of two 2D Gaussians as described by

\begin{equation}
	A_1 \times e^{-((b_1-x)/c_1)^2/2} + A_2 \times e^{-((b_2-x)/c_2)^2/2}
	\label{eq:s3}
\end{equation}

 and $I$ was normalized based on these fit parameters, so that the transmission response became either 1 or 0 depending on the parity state. Then, the normalized response was binned into two bins, 1 and $-1$, using a threshold of 0.5 and a moving window of 20 data points [visualized with red and blue lines in Fig.~\ref{fig5}(c)]. Subsequently, the amount and length of the binned data sections were counted, resulting in the amount of parity switches and the total time spent in each parity state, respectively. Finally, the parity switching rates were determined by dividing the amount of parity switches by the total time spent in each parity state. The error in the parity switching rates was found by applying bootstrapping to the normalized $I$ response with 10000 bootstrap samples.

\begin{figure}
	\includegraphics[width=0.45\textwidth]{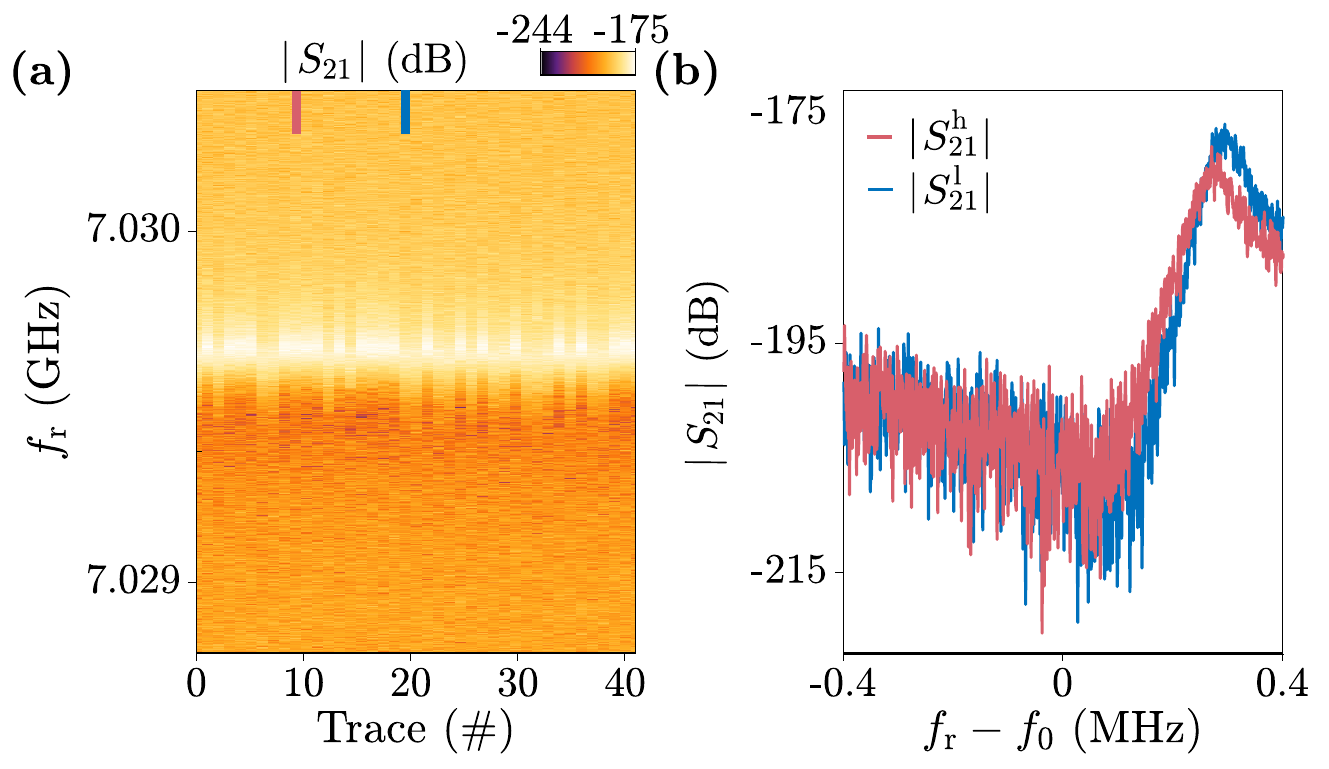}
	\caption{(a) Multiple traces of $\abs{S_{21}}$ as a function of readout frequency $f_\mathrm{r}$, acquired within a time interval of approximately 7~minutes (8~ms integration time). (b) Line cuts from (a) at the positions indicated by the markers. For the state corresponding to the response shown in red, the resonator response is less pronounced and the peak is shifted to lower frequencies with respect to the response shown in blue.}
	\label{fig9}
\end{figure}

The two traces of $\abs{S_{21}}$ as a function of readout frequency $f_\mathrm{r}$ shown in Fig.~\ref{fig5}(b) are the averaged of 19 scans acquired with an integration time of 8~ms [see Fig.~\ref{fig9}(a)]. They were measured between time traces for which the readout power was varied, resulting in a time interval of approximately 7~minutes between each scan. Here, the readout power was set to $P_\mathrm{in}=-39.1$~dBm. Due to parity switching, the resonator response traces fluctuated between two states, which had maxima offset by $\sim24$~kHz [Fig.~\ref{fig9}(b)].

\section*{APPENDIX F: SNR ESTIMATION}
\begin{figure}
	\includegraphics[width=0.3\textwidth]{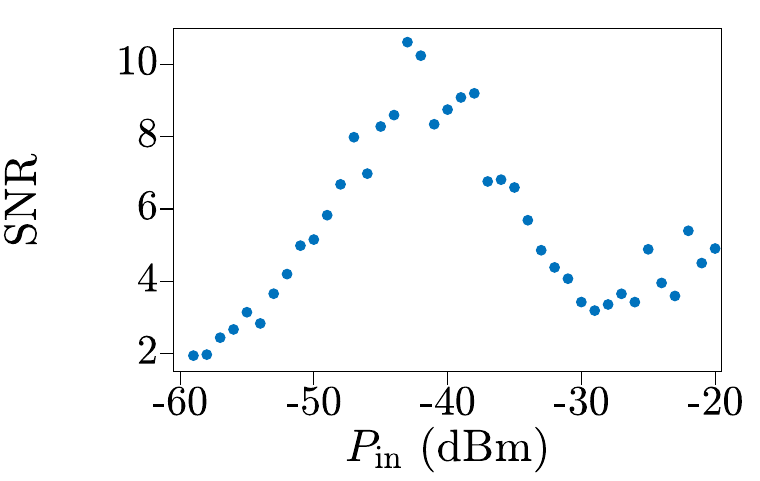}
	\caption{Signal-to-noise ratio (SNR) as a function of readout power $P_\mathrm{in}$, extracted from 400~s time traces (8~ms integration time) measured with resonator compensation at each readout power. The SNR reaches a maximum at $P_\mathrm{in}\in\{-44,-38\}$~dBm. At readout powers higher than $-30$~dBm, the SNR starts increase, likely due to the resonator entering the non-linear regime.}
	\label{fig10}
\end{figure}

Figure~\ref{fig10} depicts the signal-to-noise ratio (SNR) as a function of readout power. For each readout power, a resonator scan (at fixed power $P_\mathrm{in}=-39.1$~dBm) was taken to determine the frequency at which the subsequent time trace would be measured. This time trace consisted of 50001 data points recorded with an integration time of 8~ms, at $V_\mathrm{TG}=-731.1$~mV. The procedure for extracting the SNR is found in Ref.~\cite{hinderling2023parity}. The optimum readout power lies in the interval $P_\mathrm{in}\in\{-44,-38\}$~dBm, where the SNR shows a maximum. At readout powers higher than $-30$~dBm, the SNR increases after an initial decrease, which is indicative of the resonator being excited to the non-linear regime~\cite{Reed2010fidelity}. 

\clearpage
\newpage

\section*{Acknowledgments}
We thank the Cleanroom Operations Team of the Binnig and Rohrer Nanotechnology Center (BRNC) for their help and support. We thank F. Schupp, A. Fuhrer, W. Riess and H. Riel for useful discussions. F.~N. acknowledges support from the European Research Council (Grant~No.~804273) and the Swiss National Science Foundation (Grant~No.~200021~201082). 

\section*{Data availability}
Data presented in this work will be available on Zenodo. The data that support the findings of this study are available upon reasonable request from the corresponding author. 

\bibliography{bibliography.bib}

\newpage
\setcounter{section}{0}
\onecolumngrid

\newcounter{myc}
\renewcommand{\thefigure}{S.\arabic{myc}}

\newcounter{mye}
\renewcommand{\theequation}{S.\arabic{mye}}

\newpage	
\setlength{\parskip}{0pt}

\end{document}